\documentclass[11pt,prb]{revtex4}   
\pdfoutput=1

\usepackage{amsmath}    

\usepackage{graphicx}   

\usepackage[caption=false]{subfig}
\begin{document}

\title{Dielectric susceptibility of magnetoelectric thin films with vortex-antivortex dipole pairs}
\author{P.I.\ Karpov}
  \email{karpov.petr@gmail.com}   
\author{S.I.\ Mukhin}
  \email{sergeimoscow@online.ru}   

\affiliation{Department of Theoretical Physics and Quantum Technologies, National University of Science and Technology УMISiSФ,  Leninski avenue 4, 119049, Moscow, Russia}
\date{\today}

\begin{abstract}
We consider model of quasi-2D magnetoelectric material as XY model for spin system on a lattice with local multiferroic-like interaction of spin and electric polarization vectors. We calculate the contribution of magnetic (spin) vortex-antivortex pairs (which form electric dipoles) to the dielectric susceptibility of the system. We show that in approximation of non-interacting pairs at $T \rightarrow T_{BKT}$ (Berezinskii-Kosterlitz-Thouless temperature) dielectric susceptibility diverges.
\end{abstract}

\maketitle


\section{Introduction}

Magnetic and electric phenomena are finding more and more technological applications now.
Therefore, materials that combine properties of ferromagnets and ferroelectrics (called multiferroics) and materials that have coupled magnetic and electric subsystems (called magnetoelectrics) are interesting both
from theoretical point of view and from point of view of possible technological applications (for reviews see \cite{zvezdinufn,khom09review,mostovoy07review}).
Particulary interesting is to consider magnetoelectric and multiferroic thin films \cite{martin, wu} that contain topological defects \cite{pyatakov12JMMM, msu}.

In the present work we consider magnetoelectric material, for example type-II multiferroic (according to classification, given in \cite{khom09review}), where effective interaction of electric and magnetic subsystems leads to electric polarization (${\mathbf P}$)
and magnetization (${\mathbf M}$) coupling.
We consider an easy-plane thin film (film plane coincides with the easy plane) with quite strong interplane
interaction, so we assume that
${\mathbf P}$ and ${\mathbf M}$ vectors lie in the easy plane and their distributions are identical in all layers.
We describe the magnetic subsystem of multiferroic in terms of classical two dimensional XY model for spin system on the lattice.
Below $T_{BKT}$ existence of a single vortex is energetically unfavorable, so that all vortices are linked
in vortex-antivortex pairs \cite{kt1}. Coupling between polarization and magnetization brings us to the fact that the vortices in the
XY model possess electric charges such that an electric charge is proportional to the topological charge of the vortex \cite{mostovoy}.
The resulting dipole gas contributes to the dielectric susceptibility of the material. In this paper we calculate this contribution
of magnetic vortex-antivortex pairs to the dielectric susceptibility of the magnetoelectric material.


\section{Two dimensional XY model}

For magnetic subsystem of multiferroic we use classical two-dimensional XY model on the square lattice. Let ${\mathbf M_i} = M (\cos\phi_i, \sin\phi_i)$ be magnetic moment at $i$-th site. Then the Hamiltonian of the system can be written as
\begin{equation}
\label{hamiltonianXY}
H = - J \sum_{\langle i,j \rangle} {\mathbf M}_i {\mathbf M}_j = -J M^2 \sum_{\langle i,j \rangle} \cos(\phi_i-\phi_j).
\end{equation}
\noindent Here $\langle i,j \rangle$ denotes summation over all nearest neighbors and we consider ferromagnetic case
when exchange integral $J>0$. Taking continuous limit $\phi_i \rightarrow \phi({\mathbf x})$ we get
\begin{equation}
\label{energyXY}
H =  \frac{1}{2} \rho_s \int d^2 x (\nabla \phi)^2,
\end{equation}
\noindent where $\rho_s = J M^2$ is spin-wave stiffness and constant term is neglected.

Apart from the ground state solution $\phi = const$, there exist metastable vortex solutions
$\phi = n \arctan \frac{y-Y_v}{x-X_v} + \phi_0$.
It can be shown that energy of single vortex with winding number $n$ logarithmically diverges \cite{chaikin}: $E_v = \pi n^2 \rho_s \ln \frac{R}{a}$ ($R$ is of order of the size of the system, $a$ is the lattice spacing), therefore single vortices don't appear in macroscopic systems.

There also exist metastable configurations, which are superpositions of single-vortex solutions, for example, vortex-antivortex pair configuration
$\phi = n (\arctan \frac{y-Y_v}{x-X_v} - \arctan \frac{y-Y_a}{x-X_a})$.
Energy of such vortex-antivortex configuration is finite:
\begin{equation}
\label{Eva}
E_{va} = 2\pi n^2 \rho_s \ln\frac{X}{a}.
\end{equation}
Here $X = \sqrt{(X_v-X_a)^2 + (Y_v-Y_a)^2} \,$ is the distance between vortex and antivortex cores.

We see that single vortices don't appear in a macroscopic system because of logarithmic divergence of their energy, but system of linked vortices and antivortices has finite energy.  At temperatures $T < T_{BKT}$ when vortices and antivortices are linked into pairs, there exists some equilibrium concentration of these pairs.


\section{Including magnetoelectric coupling in XY model}

Consider the difference between pure XY model and XY model with coupled electric and magnetic subsystems. In  case of cubic lattice symmetry, keeping only the lowest-order terms,
the energy density of magnetoelectric material in electric field ${\mathbf E}$ is given by \cite{landau,mostovoy}
%
\begin{equation}
\label{energy-density}
w =  \frac{P^2}{2 \chi_e} - \gamma{\mathbf P}(({\mathbf M}\nabla){\mathbf M} - {\mathbf M}(\nabla{\mathbf M})) - {\mathbf E}{\mathbf P}+
\alpha \sum_{i,j=x,y} \partial_i M_j \cdot \partial_i M_j,
\end{equation}
\noindent where $\chi_e$ is the dielectric susceptibility in the absence of ${\mathbf M}$, $\gamma$ is the coupling constant,
and $|{\mathbf M}|=M_0=const$ is the saturation magnetization. Minimization of (\ref{energy-density}) with respect to $\mathbf P$ gives
\begin{equation}
\label{PM}
\mathbf P = \gamma\chi_e (({\mathbf M}\nabla){\mathbf M} - {\mathbf M}(\nabla{\mathbf M})) + \chi_e {\mathbf E}.
\end{equation}
\noindent Let ${\mathbf M} (\mathbf{x})=M_0 (\cos \phi(\mathbf{x}), \sin \phi(\mathbf{x}))$.
Then polarization is given by
\begin{equation}
\label{polarization}
{\mathbf P} = \gamma\chi_e M_0^2  \left(
\begin{array}{ccc}
- \partial_y\phi \\
  \partial_x\phi \\
 \end{array}
\right) + \chi_e {\mathbf E}.
\end{equation}
Inserting (\ref{PM}) to (\ref{energy-density}), we see that electric,
magnetic, and magnetoelectric parts of energy combine to
\begin{equation}
\label{energy1}
w = \left(\alpha M_0^2 - \tfrac{1}{2} \chi_e \gamma^2 M_0^4 \right) (\nabla\phi)^2 -
\chi_e \gamma M_0^2 \left(\partial_x \phi E_y - \partial_y \phi E_x\right) -
\tfrac{1}{2} \chi_e E^2.
\end{equation}\
Term $\frac{1}{2} \chi_e E^2$ gives constant contribution to the total electric susceptibility and further won't be considered.
Energy of one layer is
\begin{equation}
\label{energy-layer}
H = a \left(\alpha M_0^2 - \tfrac{1}{2} \chi_e \gamma^2 M_0^4 \right) \int d^2 r  (\nabla\phi)^2 -
a \chi_e \gamma  M_0^2 \int d^2 r (\partial_x \phi E_y - \partial_y \phi E_x).
\end{equation}

Assume, for a moment, that ${\mathbf E} = 0$. In this case expression for energy (\ref{energy-layer}) is similar to energy of one layer in XY model (\ref{energyXY})
with effective spin-wave stiffness
\begin{equation}
\label{rho}
\rho_s = (2\alpha M_0^2 - \chi_e \gamma^2 M_0^4) a.
\end{equation}
\noindent Hence, if we include the magnetoelectric coupling in conventional XY model, then in the absence of an external electric field we obtain the same XY model with different interaction constant. For typical values of parameters \cite{pyatakov12JMMM} ($\alpha M_0^2 \simeq 10^{-7}$ erg/cm, $\gamma M_0^2 \simeq 10^{-6}$ (erg/cm)$^{1/2}$, $\chi_e = 1-10$): $\alpha M_0^2$ is greater than $\gamma^2\chi_e M_0^2$ by several orders of magnitude; therefore, $\rho_s$ remains almost unchanged by magnetoelectric interaction $\rho_s \approx 2\alpha M_0^2 a$.

Let us discuss how magnetoelectric coupling affects the magnetic vortices. From (\ref{polarization}) it follows that polarization of magnetic vortex is
${\mathbf P} = -n \gamma\chi_e M_0^2 {\mathbf r}/r^2$.
As it was shown in \cite{mostovoy} this leads to appearance of electric charge in the vortex core:
\begin{equation}
\label{qe}
q_e = 2\pi n\gamma\chi_e M_0^2,
\end{equation}
\noindent where $q_e$ is a vortex charge per unit film thickness.

Further, consider what happens in non-zero electric field.
The second term on the right-hand side of (\ref{energy-layer}) becomes non-zero and ground state is reached when $\nabla\phi = \frac{\chi_e \gamma}{\rho_s} (E_y, -E_x)$. This configuration is a spin wave with period $\frac{2 \pi \rho_s}{\chi_e \gamma E}$ and it is perpendicular to the field. We see that $\lambda \rightarrow \infty$ as $E \rightarrow 0$, hence vortex pairs of finite size aren't influenced by infinitesimal field. But even at $T=0$ spin wave creates polarization ${\mathbf P} = \frac{\chi_e^2 \gamma^2 M_0^2 }{\rho_s} {\mathbf E}$ and gives contribution to susceptibility: $\chi_{spin wave} = \frac{\chi_e^2 \gamma^2 M_0^2 }{\rho_s}$. However, we expect that vortex pairs give divergent at $T_{BKT}$ contribution to susceptibility and overcome nearly constant contribution of spin waves. Further, we don't consider spin-wave contribution to susceptibility.

When we apply an external electric field, boundary effects become important. Consider single vortex, which has electric charge (\ref{qe}) due to magnetoelectric interaction. From the total electroneutrality it follows that the boundary of the sample also becomes charged; it
acquires charge of the same magnitude but opposite sign compared to the vortex.
In an external field, the boundary charge effectively shields the charge of the vortex,
and the value of screening depends on the geometry of the sample.
In this work we assume for simplicity that our sample is a disk with radius $R$. For such disk screening reduces effective charge of vortices exactly twice (see Appendix for further details).

We know that at any temperature below $T_{BKT}$ there exist a certain amount of thermally activated vortex-antivortex pairs.
Since vortex carries a positive charge and antivortex carries a negative charge, the pair forms an electric dipole.
In the next section we calculate the dielectric susceptibility of such dipole gas with a variable number of dipoles.


\section{Dielectric susceptibility calculation}

Consider a system of non-interacting vortex-antivortex quasi-2D dipole pairs of vortex lines (at temperature $T < T_{BKT}$), which exist in thin film with thickness $h$, in electric field. Here we assume that $h \ll R$ and, therefore, all lattice layers have almost identical distributions of ${\mathbf M}$, ${\mathbf P}$, and electric charge density.
For simplicity we consider only lowest-energy topological defects with $n=\pm 1$ topological charges. Electrostatic energy of such dipole in external electric field is
$- q'_e {\mathbf r} {\mathbf E}$ (where ${\mathbf r}$ is the distance between vortex and antivortex cores and $q'_e = q_e/2$, see Appendix for details).
Distance ${\mathbf r}$ can vary and this gives the contribution of vortex-antivortex interaction energy (\ref{Eva}) with effective spin-wave stiffness (\ref{rho}) to the total energy.
 Hence for the total energy per unit film thickness of such a dipole we have \cite{kt1}
\begin{equation}
\label{energy}
U_{total} =  - q'_e {\mathbf r}{\mathbf E} + 2q_m^2 \ln \frac{r}{a} - \mu,
\end{equation}
where $q_m^2 = \pi n^2 \rho_s/a$; $\mu$ is minus dipole formation energy per unit film thickness (i.e., $-\mu h$ is an energy of a dipole with charges on neighboring sites). We consider the case of low dipole concentration, therefore, $|\mu|h$ should be sufficiently large compared to $T_{BKT}$, which holds for XY model with very good accuracy \cite{kt1}. Let $U({\mathbf r}) =  - q'_e {\mathbf r}{\mathbf E} + 2q_m^2 \ln \frac{r}{a}$.
Then the grand partition function is (in units $k_B=1$;  $\beta = \frac{1}{T}$)
\begin{equation}
\label{grand0}
\begin{split}
&\mathcal{Z}({\mathbf E}, T) =  \sum_{n} \exp\left(\frac{n \mu h}{T}\right) \sum_{{\scriptstyle{different \atop configurations}}} \exp\left( -h \frac{U({\mathbf r_1})+... + U({\mathbf r_n})}{T} \right) =  \\
&=\sum_{n} \frac{1}{(n!)^2}\left(\exp(\beta\mu h)\int \frac{d^2 Q_{CM}}{a^2} \int\frac{d^2 r}{a^2} \exp\left( -\beta h U({\mathbf r}) \right)\right)^n.
\end{split}
\end{equation}
\noindent Here we replaced summation over all vortex configurations by integration: $Q_{CM}$ is the coordinate of center of mass of a dipole. Let $N$ be a number of lattice sites in one layer, then
$\exp(\beta\mu h)\int \frac{d^2 Q_{CM}}{a^2} = N \exp(\beta\mu h)$ and
\begin{equation}
\label{grand}
\begin{split}
&\mathcal{Z}({\mathbf E}, \beta) =
I_0\left( 2 \sqrt{N\exp(\beta\mu h) \int\frac{d^2 r}{a^2} \exp\left( -\beta h U({\mathbf r}) \right)} \right)= \\
&= I_0 (2\sqrt{Z_1 ({\mathbf E}, \beta)}),\\
\end{split}
\end{equation}
\noindent where $I_0(z) = \sum_{n=0}^{\infty} \frac{1}{(n!)^2} (\frac14 z^2)^n$ is the modified Bessel function of the first \cite{abramowitz} kind and $Z_1({\mathbf E}, \beta)$
is a partition function of one dipole:
\begin{equation}
\label{Z1}
Z_1(E,\beta) = N\exp(\beta\mu h)  \int\frac{d^2 r}{a^2} \exp\left( -\beta h U({\mathbf r}) \right) =\\
\frac{N\exp(\beta\mu h)}{a} \int dr \int d\varphi \left( \frac{r}{a}\right)^{-2\beta h q_m^2+1}
\exp\left(\beta h q'_e r E \cos\varphi\right).
\end{equation}
Formally, this integral diverges at large $r$. However, we should keep in mind that we are calculating linear response (i.e., we take the limit ${\mathbf E} \rightarrow 0$), so that $r$-divergence is cut off at radius of the sample $r=R$. Therefore, if we first take derivative of (\ref{Z1}) with respect to $E$ and after that let $E=0$, then we obtain the correct result.

Now, using (\ref{grand}), (\ref{Z1}), in approximation of weak field ${\mathbf E} \approx 0$, we
can calculate contribution of vortex-antivortex pairs to dielectric susceptibility of the system:
\begin{equation}
\label{chi0}
\chi_e^{(v)} = \left.\frac{T}{S h} \frac{\partial^2}{\partial E^2} \ln \mathcal{Z} \right|_{E=0} = \frac{\pi a^2 {q'}_e^2 h^2}{4 S h} \frac{N\exp(\beta\mu h) I_1 (2\delta)}{(h q_m^2/2-T) \delta \cdot I_0(2\delta)}=
\frac{\sqrt{\pi N} a^2 {q'}_e^2}{4 S} \frac{\exp (\beta\mu h/2) \sqrt{\beta h q_m^2 -1} \cdot I_1 (2\delta)}
{( h q_m^2/2-T) \cdot I_0(2\delta)},
\end{equation}
\noindent where $S$ is area of the system, $\delta = \sqrt{Z_1(E=0, \beta)}= \sqrt{\frac{\pi N \exp(\beta\mu h)}{\beta h q_m^2-1}}$.
Temperature dependence $\chi_e^{(v)}(T)$ is sketched in Fig. 1.
\begin{figure}
\includegraphics[width=0.8\textwidth]{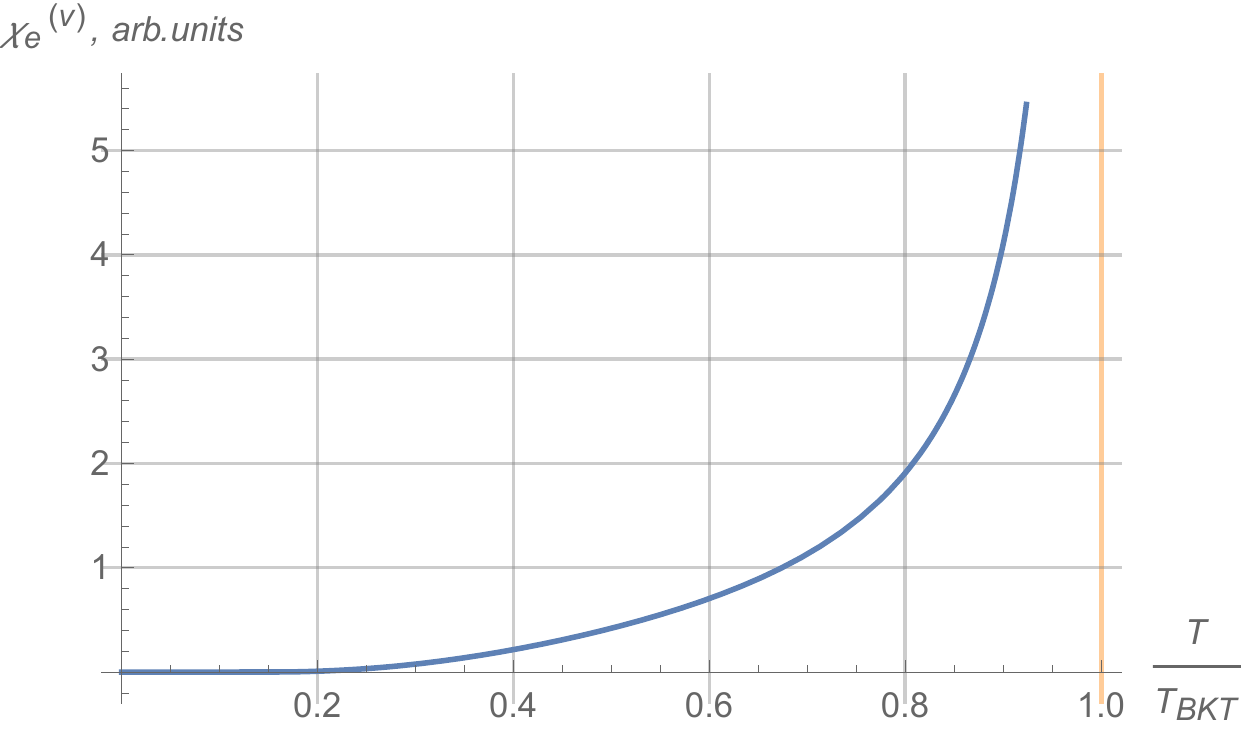}
\label{plot}
\caption{Temperature dependence of vortex contribution into dielectric susceptibility $\chi_e^{(v)}(T)$ at $T<T_{BKT}$.}
\end{figure}

Let us analyze applicability of formula (\ref{chi0}) and find it's asymptotics.
When $T \rightarrow T_{BKT} \sim h q_m^2/2$ susceptibility behaves as $\chi \simeq \frac{1}{|T-T_{BKT}|}$. This dependence (and formula (\ref{chi0})) works up to temperatures $T$, at which vortex-antivortex pairs gas can be considered as dilute, namely when the average distance between pairs $d = a \sqrt{\beta h q_m^2 - 1} \exp (\beta |\mu| h/2)$ is much greater than the average size of a pair \cite{kt1} $\langle X \rangle = a \frac{\sqrt{\beta h q_m^2 - 1}}{\sqrt{\beta h q_m^2 - 2}}$. This condition is satisfied in temperature region, where $|T-T_{BKT}| \gg \frac{1}{2} T_{BKT} \exp(-\frac{|\mu| h}{T_{BKT}})$.
Note that \cite{kt1} $\mu \simeq -\pi^2 \rho_s/a$, so from (\ref{rho}) $\mu \simeq -\pi q_m^2$.
Hence, $\frac{|\mu| h}{T_{BKT}} \simeq \frac{\pi q_m^2 h}{h q_m^2/2} = 2\pi$.
Therefore temperature region $|T-T_{BKT}| \simeq \frac{1}{2} T_{BKT} \exp(-\frac{|\mu| h}{T_{BKT}}) \approx 10^{-3} T_{BKT}$ is a narrow region near $T_{BKT}$. This means that formula (\ref{chi0}) works for all $T<T_{BKT}$, that are not very close to $T_{BKT}$.

Let us find an asymptotic behavior of $\chi_e^{(v)}$ at low temperatures.
 In this approximation $T \ll T_{BKT} \sim \frac{h q_m^2}{2}$  (which is equivalent to $\beta h q_m^2 \gg 2$) and also $\beta |\mu| h \gg 1$
(hence,   $\delta \ll 1$). Using asymptotic form $I_{\nu}(z) \sim (\frac{z}{2})^{\nu}/ \Gamma(\nu+1)$
for the modified Bessel function \cite{abramowitz},
we obtain in zeroth approximation on $E$, retaining the principal $T$-dependent term:
\begin{equation}
\label{chiapprox}
\chi_e^{(v)} = \frac{1}{2}\pi \exp(\beta\mu h) \frac{{q'}_e^2}{q_m^2} =
\frac{\pi}{2}\exp(\beta\mu h) \frac{\pi\chi_e^2\gamma^2 M_0^2}{2\alpha - \gamma^2 \chi_e M_0^2}=
\exp\left(\frac{\mu h}{T}\right) \frac{\pi^2\chi_e^2\gamma^2 M_0^2}{4\alpha - 2\gamma^2 \chi_e M_0^2}
\end{equation}
%
%
%

\section{Conclusions}

In conclusion, in this paper we investigated properties of magnetoelectric thin film with easy-plane type of symmetry and type-II multiferroic-like interaction between electric and magnetic subsystems. Magnetic vortices in such magnetoelectric possess electric charges and vortex-antivortex pairs form electric dipoles. Such dipole pairs have finite energy, so at any temperature $T<T_{BKT}$ there exists a certain amount of thermally activated vortex-antivortex pairs. We calculated the vortex-antivortex pairs contribution   in static dielectric susceptibility of the system at temperatures $T<T_{BKT}$
in approximation of non-interacting dipoles. This approximation is valid at temperatures which are not extremely close to $T_{BKT}$ (when $|T-T_{BKT}| \gg \frac{1}{2} T_{BKT} \exp(-\frac{|\mu| h}{T_{BKT}})$).
In the low temperature limit dielectric susceptibility (\ref{chi0}) behaves as an activation exponential (\ref{chiapprox}), which is consistent with the fact that at $T\rightarrow 0$ number of vortex pairs is proportional to $\exp\left(\frac{\mu h}{T}\right)$. As $T \rightarrow T_{BKT}$ formula (\ref{chi0}) gives diverging susceptibility. This reflects the process of vortex-antivortex pairs unbinding and the phase transition. However, we expect that close to $T_{BKT}$ interaction of dipole pairs becomes important and, therefore, at $T_{BKT}$ susceptibility stays finite.


\begin{acknowledgements}
 Authors are grateful to Igor S. Burmistrov, Daniel I. Khomskii and Maxim V. Mostovoy for helpful discussions. PK acknowledges financial support by the non-profit Dynasty foundation.
\end{acknowledgements}

\appendix*

\section{}
\begin{figure}
\subfloat[Distribution of polarization for single vortex. \label{disk}]{%
  \includegraphics[width=.45\linewidth]{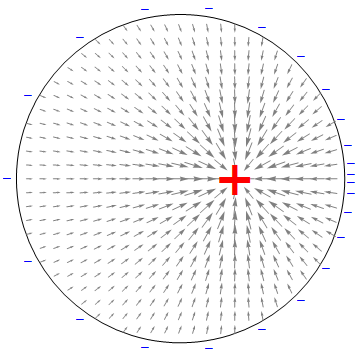}%
}
\hfill
\subfloat[Coordinate system. \label{coord}]{%
  \includegraphics[,width=.51\linewidth]{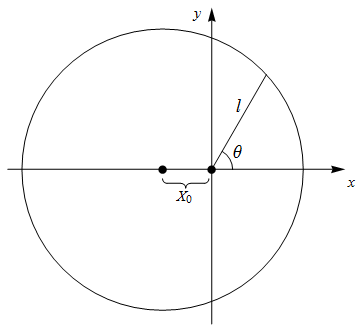}%
}
\caption{(a) Distribution of polarization for single vortex. Vortex core is positively charged, edge of the sample is negatively charged.
         (b) Coordinate system. Vortex core is placed at $(0,0)$; disk center is $(-X_0,0)$. }
\label{figappendix}
\end{figure}

In this Appendix we calculate energy of charged vortex in electric field. As it was explained in the main text, magnetic vortex core acquires electric charge (\ref{qe}) due to magnetoelectric coupling (\ref{energy-density}). Since our sample is electrically neutral, it's edge acquires negative charge of the same magnitude, which should be taken into account (see Fig. \ref{disk}).

Let our sample be disk with radius $R$, which contains single vortex with vorticity $n$, placed at distance $X_0$ from the center of the disk. Choose а coordinate system as shown in Fig. \ref{coord}; electric field ${\mathbf E} = (E_x, E_y)$.

From (\ref{polarization}) polarization of single vortex configuration is
\begin{equation}
\label{P-app}
{\mathbf P} = \gamma\chi_e M_0^2  \left(
\begin{array}{ccc}
- \partial_y\phi \\
  \partial_x\phi \\
 \end{array}
\right) =
-\frac{n\gamma\chi_e M_0^2}{r}  \left(
\begin{array}{ccc}
  \cos\theta \\
  \sin\theta \\
 \end{array}
\right).
\end{equation}
Since we calculate linear response, $\chi_e {\mathbf E}$ term in (\ref{polarization}) was omitted.
Then electric energy of vortex configuration per unit film thickness is
\begin{equation}
U_{el} = - \int {\mathbf P} {\mathbf E} \, d^2 r = n\gamma\chi_e M_0^2 \int_0^{2\pi} d\theta \, \int_0^{l(\theta)} dr \, (E_x \cos\theta + E_y \sin\theta) = n\gamma\chi_e M_0^2 \int_0^{2\pi} d\theta  (E_x \cos\theta + E_y \sin\theta) l(\theta).
\end{equation}
Since $l(\theta) = \sqrt{R^2 - X_0^2 \sin^2\theta} -X_0 \cos\theta \,$ is an even function of $\theta$, then odd term $E_y\sin\theta$ vanishes. Thus,

\begin{equation}
U_{el} =  2 n\gamma\chi_e M_0^2 E_x \int_0^{\pi} d\theta  \cos\theta \left(\sqrt{R^2 - X_0^2 \sin^2\theta} -X_0 \cos\theta \right) = -\pi n\gamma\chi_e M_0^2 E_x X_0.
\end{equation}

Using (\ref{qe}) we see that $U_{el} = -\frac{1}{2} q_e  E_x X_0$ or for general displacement of vortex core ${\mathbf r} = (X_0, Y_0)$
\begin{equation}
\label{energy-app}
U_{el} = -q'_e  {\mathbf E} {\mathbf r},
\end{equation}
where $q'_e = q_e/2$ is an effective electric charge of the vortex, which is useful for finding electric energy of the vortex in an external electric field.

Since polarization (\ref{P-app}) is linear in phase angle $\phi$, the same result (\ref{energy-app}) holds for vortex-antivortex configuration, with ${\mathbf r}$ denoting a vector that connects cores of the antivortex and the vortex.

\end{document}